\documentclass[12pt]{article}
\usepackage[english,german,french,polish]{babel}
\usepackage[T1]{fontenc}
\usepackage{amsfonts}

\begin{document}

\selectlanguage{english}

\baselineskip 0.77cm
\topmargin -0.6in
\oddsidemargin -0.1in

\let\ni=\noindent

\renewcommand{\thefootnote}{\fnsymbol{footnote}}

\newcommand{\SM}{Standard Model }

\pagestyle {plain}

\setcounter{page}{1}

\pagestyle{empty}

~~~

\begin{flushright}
IFT-- 06/15
\end{flushright}

\vspace{0.4cm}

{\large\centerline{\bf New structure for the off-diagonal part }}

{\large\centerline{\bf of neutrino effective mass matrix{\footnote{Work supported in part by the Polish Ministry of Higher Education and Science, grant 1 PO3B 099 29 (2005-2007). }}}} 

\vspace{0.4cm}

{\centerline {\sc Wojciech Kr\'{o}likowski}}

\vspace{0.3cm}

{\centerline {\it Institute of Theoretical Physics, Warsaw University}}
 
{\centerline {\it Ho\.{z}a 69,~~PL--00--681 Warszawa, ~Poland}}

\vspace{0.5cm}

{\centerline{\bf Abstract}}

\vspace{0.2cm}

Two $3\times 3$ matrices playing the role of annihilation and creation operators in the 
neutrino-flavor space of $\nu_e , \nu_\mu , \nu_\tau$ are applied to construct the off-diagonal
part of neutrino effective mass matrix $M$. The construction leads to a new relation between two 
off-diagonal elements $M_{e\,\mu}$ and $M_{\mu\,\tau}$ of $M$. In the case of tribimaximal 
neutrino mixing, this relation implies a new mass sum rule for $\nu_1 , \nu_2 , \nu_3$: $m_3  
= \eta \,m_2 - (\eta - 1)m_1$. In a plausible option consistent with the charged-lepton mass spectrum,
one gets $\eta \equiv (2/3)(4\sqrt{3}+1) = 5.28547$. Then, ${\Delta m^2_{32}}/{\Delta m^2_{21}} = \eta^2 -1 = 26.9362$ if $m_1 = 0$, what {\it predicts} $\Delta m^2_{32} \sim 2.2\times 10^{-3}\;{\rm eV}^2$ when the input $\Delta m^2_{21} \sim 8.0\times 10^{-5}\;{\rm eV}^2$ is used.

\vspace{0.5cm}

\ni PACS numbers: 12.15.Ff , 14.60.Pq  .

\vspace{0.6cm}

\ni August 2006  

\vfill\eject

~~~
\pagestyle {plain}

\setcounter{page}{1}

\vspace{0.3cm}

\ni {\bf 1. Previous work}

\vspace{0.3cm}

In a recent paper [1], we have proposed the new relation
  
\begin{equation}
M_{\mu\,\tau} = 4\sqrt3 \,M_{e\,\mu} 
\end{equation}

\ni between two off-diagonal elements of neutrino effective neutrino mass-matrix $M = \left( M_{\alpha\,\beta} \right) \;\;(\alpha , \beta = e, \mu, \tau)$. This relation is valid, if the off-diagonal part of $M$ is tentatively conjectured to be built up from two $3\times 3$ matrices playing the role of annihilation and creation operators acting in the neutrino-flavor space of $\nu_e , \nu_\mu , \nu_\tau$ and, in addition, if the neutrino flavor-weighting factors $\rho_\alpha \;(\alpha = e , \mu , \tau)$ defined in Eq. (12) get the specific values

\vspace{-0.1cm}

\begin{equation}
\rho_e = \frac{1}{29} \;,\; \rho_\mu = \frac{4}{29} \;,\; \rho_\tau = \frac{24}{29}\,,
\end{equation}

\vspace{0.2cm}

\ni where $\sum_\alpha \rho_\alpha = 1$. In Ref. [1], the factors $\rho_\alpha$ have been tentatively assumed to be numerically equal to the fermion generation-weighting factors $\rho_i\; (i = 1,2,3)$  which take the specific values (2), $\rho_1 = 1/29, \rho_2 = 4/29, \rho_3 = 24/29$, for all fundamental fermions and play a significant role in describing fermion mass spectra [2,3,4]. Especially, appearing  in an efficient charged-lepton empirical mass formula, the factors $\rho_i$ are consistent with the precisely measured mass spectrum $m_e, m_\mu, m_\tau $ [3,4]. Consequently, in Ref. [1], the annihilation and creation operators in the neutrino-generation space of $\nu_1 , \nu_2 , \nu_3$ have been applied, rather than those having the same matrix form in the neutrino-flavor space of $\nu_e , \nu_\mu , \nu_\tau$ (see the comment after Eq. (12)).

In the case when the neutrino effective mixing matrix $U = \left( U_{\alpha\,i} \right) \;\;(\alpha = e, \mu, \tau\,,\, i=1,2,3)$ is of the tribimaximal form [5]

\begin{equation}
U = \left( \begin{array}{rrr} \frac{\sqrt2}{\sqrt3} & \frac{1}{\sqrt3} & 0\, \\ - \frac{1}{\sqrt6} & \frac{1}{\sqrt3} & \frac{1}{\sqrt2} \\ \frac{1}{\sqrt6} & -\frac{1}{\sqrt3} & \frac{1}{\sqrt2}  \end{array} \right) 
\end{equation}

\vspace{0.2cm}

\ni that is reasonably consistent with all confirmed neutrino experiments, our relation (1) provides the new mass sum rule for $\nu_1 , \nu_2 , \nu_3$ 
 
\begin{equation}
m_3  =  \eta \,m_2 - (\eta - 1) \,m_1 
\end{equation}
 
\ni with the specific coefficient
 
\begin{equation}
\eta \equiv \frac{2}{3} \,(4\sqrt3 +1) = 5.28547\,.
\end{equation}

\ni In fact, in the tribimaximal case, the elements $M_{\alpha\,\beta} = \sum_i U_{\alpha i}m_i U^*_{\beta i}$ of the netrino effective mass matrix $M$ are:

\begin{eqnarray}
M_{e\,e} & = & \:\:\:\frac{1}{3}(2m_1 + m_2)\,, \nonumber \\
M_{\mu\,\mu} & = & \:\:\,M_{\tau\,\tau} = \:\:\:\,\frac{1}{6} (m_1 + 2m_2 + 3 m_3)\,, \nonumber \\
M_{e\,\mu} & = & -M_{e\,\tau} =  -\frac{1}{3} (m_1 - m_2)\,, \nonumber \\
M_{\mu\,\tau} & = & -\frac{1}{6} (m_1 + 2m_2 - 3 m_3)\,.
\end{eqnarray}

\ni Two last equations (6), if inserted in the relation (1), give the mass sum rule (4) with (5). 

If $m_1 = 0$, the mass sum rule (4) with (5) leads to the {\it prediction}

\begin{equation}
{\Delta m^2_{32}}/{\Delta m^2_{21}} = \eta^2 -1 = 26.9362
\end{equation}

\ni and hence, $\Delta m^2_{32} \sim 2.2\times 10^{-3}\;{\rm eV}^2$, when the input of experimental estimate $\Delta m^2_{21} \sim 8.0\times 10^{-5}\;{\rm eV}^2$ [5] is used. The predicted $\Delta m^2_{32}$ is close to the popular experimental best fit $\Delta m^2_{32} \sim 2.4\times 10^{-3}\;{\rm eV}^2$ [6].

\vspace{0.3cm}

\ni {\bf 2. A less restrictive case}

\vspace{0.3cm}

In the present paper, we find the counterpart of formula (1) in the case when the neutrino 
flavor-weighting factors $\rho_\alpha \;(\alpha = e, \mu, \tau)$ are not necessarily of the form (2) characteristic for the fermion generation-weighting factors $\rho_i \;(i=1,2,3)$, and where the neutrino mixing may deviate from the tribimaximal pattern (3).

To this end, let us make use of two matrices in the neutrino-flavor space of $\nu_e , \nu_\mu , \nu_\tau$:

\begin{equation} 
a = \left( \begin{array}{ccc} 0 & 1 & 0 \\ 0 & 0 & \sqrt2 \\ 0 & 0 & 0 \end{array}\right)  \; ,\;a^\dagger = \left( \begin{array}{ccc} 0 & 0 & 0 \\ 1 & 0 & 0 \\ 0 & \sqrt2 & 0 \end{array}\right)\;,  
\end{equation}

\ni satisfying the relations

\begin{equation} 
n = a^\dagger a \;,\; [a\, , \,n] = a \;,\; [a^\dagger\, , \,n] = -a^\dagger\;, \;a^3 = 0 \;,\; a^{\dagger \,3}= 0
\,.  
\end{equation}

\ni Though $[a\,,\,a^\dagger] \neq {\bf 1}$, such $a$ and $a^\dagger $ play the role of annihilation and creation operators in the neutrino-flavor space of $\nu_e , \nu_\mu , \nu_\tau$, while

\vspace{-0.1cm}

\begin{equation} 
n = \left( \begin{array}{ccc} 0 & 0 & 0 \\ 0 & 1 & 0 \\ 0 & 0 & 2 \end{array}\right)
\end{equation}

\vspace{0.1cm}

\ni is the operator defining formal occupation numbers $n_\alpha = 0,1,2 $ corresponding to three flavors $\alpha = e , \mu , \tau$ of $\nu_e , \nu_\mu , \nu_\tau$, respectively.

Then, it is natural to conjecture tentatively that for the off-diagonal part of neutrino effective mass matrix $ M $ the following constraining identity holds:

\vspace{-0.1cm}

\begin{eqnarray} 
\left( \begin{array}{rrr} 0\:\: & M_{e\,\mu} & M_{e\,\tau} \\ M_{e\,\mu}  & 0\:\: & M_{\mu\,\tau} \\ 
M_{e\,\tau}  & M_{\mu\,\tau} & 0\:\: \end{array}\right) & = & \mu \,\rho^{1/2} \left[g(a + a^\dagger) + g'(a^2 + a^{\dagger\,2})\right] \rho^{1/2} \nonumber \\ & & \nonumber \\ 
& = & \mu \,\left( \begin{array}{ccc} 0 & g\sqrt{\rho_e \rho_\mu} & g'\sqrt{2\rho_e \rho_\tau} \\ g\sqrt{\rho_e \rho_\mu} & 0 & g\sqrt{2\rho_\mu \rho_\tau} \\ g'\sqrt{2\rho_e \rho_\tau} & g\sqrt{2\rho_\mu \rho_\tau} & 0 \end{array}\right)\;\,, 
\end{eqnarray}

\vspace{0.1cm}
 

\ni where $g$ and $g'$ are free parameters (multiplied by an active-neutrino mass scale $\mu >0$), while

\vspace{-0.2cm}

\begin{equation} 
\rho^{1/2} = \left( \begin{array}{ccc} \rho^{1/2}_e & 0 & 0 \\ 0 & \rho^{1/2}_\mu & 0 \\ 0 & 0 & \rho^{1/2}_\tau \end{array}\right) \,.  
\end{equation}
 
\vspace{0.2cm}

\ni Here, $\rho_\alpha\; (\alpha = e , \mu , \tau) $ are called the neutrino flavor-weighting factors, not necessarily equal to the fermion generation-weighting factors $\rho_i (i = 1,2,3) $ (${\rm Tr} \rho \equiv \sum_\alpha \rho_\alpha = 1$). Formally, we have in Eq. (11) $ \rho^{1/2} = \left( \rho^{1/2}_{\alpha \beta}\right) =  \left( \delta_{\alpha \beta}\, \rho^{1/2}_\beta \right)$ and $ a = \left( a_{\alpha \beta}\right)$. But, in the case of $\rho_\alpha $ taking as in Ref. [1] the plausible values (2) (numerically equal to $\rho_i$), we can alternatively write $ \rho^{1/2} = \left( \rho^{1/2}_{\alpha i}\right) =  \left( \delta_{\alpha i}\, \rho^{1/2}_i \right)$ and $ a = \left( a_{i j}\right)$, the latter acting in the neutrino-generation space of $\nu_1 , \nu_2, \nu_3$ rather than in the neutrino-flavor space of $\nu_e , \nu_\mu , \nu_\tau$ (here, both $a$'s have the same matrix form (8)). Above, we put $\delta_{\alpha i} = 1$ for $\alpha\,i = e\,1 , \mu\,2 ,\tau\,3$ and = 0 otherwise. Thus, in this case, the mechanism (11) constraining the off-diagonal part of $M$ can be interpreted as working entirely in the 
neutrino-generation space, since among all {\it a priori} possible components $\rho_{\alpha i}$ of $\rho$, only those with $\alpha\, i = e\,1 , \mu\,2 ,\tau\,3$ can be nonzero.

From the identity (11), it follows that

\begin{equation}
g = M_{e\,\mu} /(\mu\, \sqrt{\rho_e \rho_\mu}) = M_{\mu\,\tau} /(\mu\, \sqrt{2\rho_\mu \rho_\tau}) \;,\; g' = M_{e\,\tau} /(\mu\, \sqrt{2\rho_e \rho_\tau}) 
\end{equation}

\ni and hence,
  
\begin{equation}
M_{\mu\,\tau} = \sqrt{2\rho_\tau/\rho_e}\, M_{e\,\mu} \;.
\end{equation}

\ni Note that $M_{e \mu} = -M_{e \tau}$ in the case of bilarge or (in particular) tribimaximal neutrino mixing. Then, from Eq. (13)

\begin{equation}
g = -\sqrt{2\rho_\tau/\rho_\mu} \;g' \,.
\end{equation}

In the tribimaximal case (see Eq. (6)), Eqs. (13) and (14) imply the parameter values 

\begin{equation}
g = -\sqrt{2\rho_\tau/\rho_\mu} \,g' = \frac{1}{3}(m_2 - m_1)/\sqrt{\rho_e\,\rho_\mu}
\end{equation}

\ni and the neutrino mass sum rule
 
\begin{equation}
m_3  =  \eta \,m_2 - (\eta - 1) \,m_1 \,
\end{equation}
 
\ni with the coefficient
 
\begin{equation}
\eta \equiv \frac{2}{3} \,(\sqrt{2\rho_\tau/\rho_e} +1) \,.
\end{equation}

In the case of $\rho_\alpha \;(\alpha = e , \mu , \tau) $ taking tha plausible values (2), we get from Eqs. (16) and (18) the parameter values $g = - 2\sqrt3\, g' = [29/(6\mu)](m_2 - m_1)$ and the coefficient $\eta \equiv (2/3)(4 \sqrt3 +1)= 5.28547 $ as given in Eq. (5).
 
\vspace{0.3cm}

\ni {\bf 3. Conclusion}

\vspace{0.3cm}

In conclusion, the constraining identity (11), if it holds for the off-diagonal part of neutrino effective mass matrix $M$, means that this part is built up from two annihilation and creation operators acting in neutrino flavor-space of $\nu_e , \nu_\mu , \nu_\tau$. Then, the new relation (14) is valid between $M_{e \mu}$ and $M_{\mu \tau}$, implying in the case of tribimaximal neutrino mixing the new neutrino mass sum rule presented in Eqs. (17) and (18). If, in addition, the flavor-weighting factors $\rho_\alpha\; (\alpha = e , \mu , \tau)$ take the plausible values (2), this mass sum rule is given as in Eqs. (4) and (5), while the relation (14) assumes the form (1).

For normal hierarchy of neutrino masses $m_1 < m_2 < m_3$, when the lowest mass is put in the range

\begin{equation}
m_1 \sim (0 \;{\rm to}\;10^{-3})\;{\rm eV} \;,
\end{equation}

\ni we infer from the experimental estimate $\Delta m^2_{21} \sim 8.0\times 10^{-5}\;{\rm eV}^2$ that
 
\begin{equation}
m_2 \sim (8.9 \;{\rm to}\; 9.0)\times10^{-3}\;{\rm eV} \;.
\end{equation}

\ni Then, making use of the mass sum rule (4) with (5), valid in the case of $\rho_\alpha$ taking the plausible values (2), we {\it predict}

\begin{equation}
m_3 \sim (4.7\;{\rm to}\; 4.3)\times 10^{-2} \;{\rm eV}
\end{equation}
 
\ni and hence,

\begin{equation} 
\Delta m^2_{32} \sim (2.2 \; {\rm to} \; 1.8)\times 10^{-3}\;{\rm eV}^2
\end{equation}

\ni for $m_1$ lying in the range (19). The value $\Delta m^2_{32} \sim 2.2\times 10^{-3}\;{\rm eV}^2$ is close to the popular experimental best fit $\Delta m^2_{32} \sim 2.4\times 10^{-3}\;{\rm eV}^2$ [6]. So, in the range (19), the value $m_1 \sim 0$ is implied. Then, $m_3 \sim \eta \,m_2$ from Eq. (17).

In the case of $m_1 = 0$, Eqs. (6) and (4) with (5) imply the relations

\begin{eqnarray}
M_{e\,e} & = & M_{e\,\mu} =  -M_{e\,\tau} = \frac{1}{3} m_2\,, \nonumber \\
M_{\mu\,\mu} & = & \:\:\,M_{\tau\,\tau} = \frac{1}{3} m_2 + \frac{1}{2} m_3 = (4\sqrt3 +2)\,\frac{1}{3}\, m_2 = 8.92820\,\frac{1}{3}\, m_2 \,, \nonumber \\
M_{\mu\,\tau} & = &   -\frac{1}{3} m_2 + \frac{1}{2} m_3 = 4\sqrt3 \,\frac{1}{3}\, m_2 = 6.92820\,\frac{1}{3} m_2\,,
\end{eqnarray}

\ni where $m_2 \sim 8.9 \times 10^{-3}\;{\rm eV}$. This leads to the proportion

\begin{equation} 
M_{e\,e} : M_{e\,\mu} :  (-M_{e\,\tau}) : M_{\mu\,\tau} : M_{\mu\,\mu} : M_{\tau\,\tau} = 1 : 1 : 1 : 4\sqrt3 : (4\sqrt3 +2) : (4\sqrt3 +2)
\end{equation}

\ni for elements of the matrix $M$. Note that Tr$M = (\eta +1) m_2 = 6.28547\, m_2$.

\vspace{0.3cm}

\ni {\bf 4. Final remark}

\vspace{0.3cm}

Note finally that the identity (11), involving the flavor-weighting factors $\rho_\alpha\; (\alpha = e , \mu , \tau)$ which take plausibly the same values as the generation-weighting factors $\rho_i\; (i =  1,2,3)$, is related in spirit with the empirical mass formula we have proposed recently [4] for active mass neutrinos $\nu_i$ of three generations $i = 1,2,3$. This formula reads

\begin{equation}
m_{i}  = \mu \, \rho_i \left[1- \frac{1}{\xi} \left(N^2_i + \frac{\varepsilon -1}{N^2_i}\right)\right] \;,
\end{equation}

\ni where $N_i = 1+2n_i = 1,3,5$ ($n_i = 0,1,2$) and $\rho_i = 1/29\,,\,4/29\,,\,24/29$, while  $\mu >0\,,\, \varepsilon $ and $\xi $ are three free parameters. Explicitly, the mass formula (25) can be rewritten as follows: 

\begin{eqnarray}
m_1 & = & \frac{\mu}{29} (1 - \frac{\varepsilon}{\xi}) \,, \nonumber \\
m_2 & = & \frac{\mu}{29}\, 4\left[1 -\frac{1}{9\xi}(80 + \varepsilon)\right] \,, \nonumber \\
m_3 & = & \frac{\mu}{29} \,24\left[ 1 -\frac{1}{25\xi}\,(624 + \varepsilon)\right] \,.
\end{eqnarray}

\ni Due to the formula (25) or (26), the mass sum rule (4) with (5), valid in the case of $\rho_\alpha $ taking the plausible values (2), imposes on two parameters $\varepsilon$ and $\xi $ the constraint

\begin{equation}
\frac{1}{\xi} = 0.0173763 - 0.0070452 \frac{\varepsilon}{\xi} \;.
\end{equation}

For $m_1 = 0$, three parameters in the neutrino mass formula (25) or (26) take the values

\begin{equation}
\mu \sim 8.1\times 10^{-2} \;{\rm eV}\;,\;  \frac{\varepsilon}{\xi} = 1 \;,\;  \frac{1}{\xi} = 1.03311\times 10^{-2} \;.
\end{equation}

\ni Here, in contrast to the value of $\mu$, the values of $\varepsilon/\xi $ and $1/\xi $ are independent of the input of experimental estimate $\Delta m^2_{21} \sim 8.0\times 10^{-5}\;{\rm eV}^2$, since they are evaluated from the relations $m_1 = 0$ and $m_3 = \eta\,m_2$, where $\mu > 0$ is cancelled after Eqs. (26) are used ($m_1 = 0$ implies $\varepsilon/\xi = 1$).

\vfill\eject

~~~~
\vspace{0.5cm}

{\centerline{\bf References}}

\vspace{0.5cm}

{\everypar={\hangindent=0.6truecm}
\parindent=0pt\frenchspacing

{\everypar={\hangindent=0.6truecm}
\parindent=0pt\frenchspacing

[1]~ W. Kr\'{o}likowski, {\tt hep--ph/0606223} (here, the notation $\rho_i$ is used also for $\rho_\alpha$  that is numerically equal to $\rho_i$).

\vspace{0.2cm}

[2]~W. Kr\'{o}likowski, {\it Acta Phys. Pol.} {\bf B 36}, 2051 (2005) [{\tt hep-ph/0503074}]; {\it Acta Phys. Pol.} {\bf B~37}, 1781 (2006) [{\tt hep-ph/0510355}].

\vspace{0.2cm}

[3]~W. Kr\'{o}likowski, {\it Acta Phys. Pol.} {\bf B 32}, 2961 (2001) [{\tt hep-ph/0108157}], Appendix; {\it Acta Phys. Pol.} {\bf B~33}, 2559 (2002) [{\tt hep-ph/0203107}]; and references therein.

\vspace{0.2cm}

[4]~W. Kr\'{o}likowski, {\tt hep--ph/0602018}; {\tt hep--ph/0604148}.

\vspace{0.2cm}

[5]~L. Wolfenstein,  {\it Phys. Rev.} {\bf D 18}, 958 (1978); P.F. Harrison, D.H. Perkins and W.G.~Scott, {\it Phys. Lett.} {\bf B 458}, 79 (1999); {\it Phys. Lett.} {\bf B 530}, 167 (2002); Z.Z.~Xing. {\it Phys. Lett.} {\bf B 533}, 85 (2002); P.F. Harrison, and W.G.~Scott, {\it Phys. Lett.} {\bf B 535}, 163 (2003); T.D.~Lee, {\tt hep--ph/0605017}.

\vspace{0.2cm}

[6]~{\it Cf. e.g.} G.L. Fogli, E. Lisi, A. Marrone and A. Palazzo, {\tt hep--ph/0506083}; T.~Schwetz {\tt 
hep--ph/0606060}; F.~Valle, {\tt hep--ph/0608101}.

\vfill\eject

\end{document}